\documentclass[prb,twocolumn,amsmath,amssymb,superscriptaddress]{revtex4}
\usepackage[T1]{fontenc}
\usepackage[latin1]{inputenc}
\usepackage{graphics}
\usepackage{amssymb}
\usepackage{amsmath}
\usepackage{overpic}

\newcommand{\be}{\begin{equation}}
\newcommand{\ee}{\end{equation}}
\newcommand{\bea}{\begin{eqnarray}}
\newcommand{\eea}{\end{eqnarray}}

\def\lb{\label}

\newcount\bozza \bozza=1
\ifnum\bozza=1
\newdimen\shift \shift=-2truecm
\def\lb#1{%
{\label{#1}\rlap{\kern\shift{$\scriptstyle#1$}}}}
\else\def\lb#1{\label{#1}} \fi

\begin{document}

\title{Extended Drude model and role of interband transitions
in the mid-infrared spectra of pnictides}

\author{L. Benfatto}

\affiliation
{Institute for Complex Systems (ISC), CNR, U.O.S. Sapienza and \\
Department of Physics, Sapienza University of Rome, P.le A. Moro 2,
00185 Rome, Italy}

\author{E. Cappelluti}

\affiliation
{Instituto de Ciencia de Materiales de Madrid,
ICMM-CSIC, Cantoblanco, E-28049 Madrid, Spain}

\affiliation
{Institute for Complex Systems (ISC), CNR, U.O.S. Sapienza and \\
Department of Physics, Sapienza University of Rome, P.le A. Moro 2,
00185 Rome, Italy}

\author{L. Ortenzi} 
\affiliation{Max-Planck-Institut f\"{u}r Festk\"{o}rperforschung,
Heisenbergstra$\mathrm{\beta}$e 1, D-70569 Stuttgart, Germany}

\author{L. Boeri}
\affiliation{Max-Planck-Institut f\"{u}r Festk\"{o}rperforschung,
Heisenbergstra$\mathrm{\beta}$e 1, D-70569 Stuttgart, Germany}
\date{\today}

\begin{abstract}
We analyze the outcomes of an extended-Drude-model approach to the
optical spectra of pnictides, where the multiband nature of the
electronic excitations requires a careful analysis
of the role of interband processes in the optical conductivity.
Through a direct comparison between model
calculations of the intraband optical spectra and experimental
data, we show that interband transitions,
whose relevance is shown by first-principle calculations,
give a non negligible
contribution already in the infrared region. This leads to a substantial
failure of the extended-Drude-model analysis on the
measured optical data without subtraction of interband
contributions.
\end{abstract}

\pacs{74.25.Gz,74.70.Xa, 74.25.nd,74.25.Jb }

\maketitle

Optical studies are a useful experimental probe to 
analyze the interactions at play in different classes of correlated
materials,~\cite{basov_rmp} for which
it would be desirable to devise a common theoretical scheme.
A typical example is  the so-called
extended-Drude-model (EDM) analysis, where the experimental 
optical conductivity is analyzed in terms of a
Drude-like model with a frequency-dependent inverse lifetime and
effective mass.\cite{basov_rmp} 
This approach is well justified 
for a single-band system interacting with a continuum of
bosonic excitations (such for instance phonons in a metal)
where it  can provide useful information on 
the relevant collective modes interacting with the
electronic particle-hole excitations.\cite{Shulga}
The EDM has  been widely used also for strongly correlated materials,
such as the cuprate superconductors;
here, intraband excitations are
well separated from the
optical interband transitions, which pose a natural cut-off
(typically of order of 4000 cm$^{-1}$) to the
applicability of the EDM itself in these compounds.\cite{Shulga,physica_c}

After the discovery of superconductivity in iron-based superconductors
an intense experimental and theoretical research has been devoted to
the EDM analysis in these materials as well.\cite{qz,yang,wu1,tu,barisic} A
general outcome of this analysis is a relatively large
frequency-dependent in-plane inverse lifetime $\tau^{-1}(\omega)\sim
\omega$,\cite{qz,yang} that, within the EDM, is interpreted in terms
of a strong-coupling regime ($\lambda\simeq 3-4$) for the relevant
bosonic excitations,\cite{yang,wu1} located around 20-60 meV. These
are usually identified with the spin fluctuations between the hole and
electron pockets of the Fermi surface, that from the very beginning
have been suggested as the most promising candidates for the
superconducting pairing.\cite{mazin}

There are however two main open issues concerning these results.  On
one hand, a strong coupling to these low-energy bosonic modes would
imply also an effective mass at low energy much larger than that
measured by other probes,\cite{cv} such as specific-heat\cite{hardy10}
or de Haas-van Alphen (dHvA).\cite{analytis10} On the other hand, the
use of the EDM implicitly assumes that the interband transitions in
pnictides become relevant only above a threshold frequency of about
$\sim 2000$ cm$^{-1}$, even though some authors suggested that they
are present at lower energies.\cite{heumen1,drechsler,boris1}
Low-energy interband
transitions may invalidate the conclusions of the EDM analysis, 
introducing spurious effects.

%
In this paper, we quantify the relative role of interband and
intraband transitions in the EDM analysis of the in-plane
optical conductivity in pnictides, focusing on
LaFePO where no magnetic transition occurs and where the
coupling to spin fluctuations within a multiband Eliashberg approach
has been already discussed in the context of different physical
properties.\cite{ortenzi,sumrule} Through a direct comparison with the
experimental optical data of Ref. [\onlinecite{qz}] we show
that, for the physical range of the coupling of electrons with
spin-fluctuations, the intraband contribution alone is not sufficient to account for the large inverse
lifetime obtained by the EDM analysis. In this context we also show that
low-energy interband transitions, individuated through first-principles
calculations, play a dominant role here already at very low energy
($\omega \gtrsim 500$ cm$^{-1}$), invalidating the conclusion of a simple
EDM analysis of the data.  Such a result, along with the similarity of
the band structures and optical spectra of Fe pnictides, suggests that
a new approach is needed in order to analyze the optical data of
pnictides.

The EDM analysis  is based on the fact that
many-body interactions modify the Drude formula for the optical
conductivity by introducing a frequency dependence into the 
inverse lifetime $\tau^{-1}(\omega)$ and a mass enhancement factor 
$m^*(\omega)/m_b$, where $m_b$ is the band mass of the carriers.\cite{basov_rmp}
Deconvoluting the real $\sigma_1(\omega)$  and imaginary
$\sigma_2(\omega)$ part of the complex optical conductivity, one can
extract:
\begin{eqnarray}
\tau^{-1}(\omega)
&=&
\frac{\omega_P^2}{4\pi}\frac{\sigma_1(\omega)}{\sigma_1^2(\omega)+\sigma_2^2(\omega)},
\label{tau}
\\
\frac{m^*(\omega)}{m}
&=&
\frac{\omega_P^2}{4\pi \omega}
\frac{\sigma_2(\omega)}{\sigma_1^2(\omega)+\sigma_2^2(\omega)},
\label{m}
\end{eqnarray}
where $\omega_P$ is the plasma frequency, that is usually estimated
integrating the real part of the conductivity up to a cut-off
$\omega_c$. The underlying idea is that the data below $\omega_c$
represent intraband optical transitions of {\em interacting} charge carriers (with
spectral weight proportional to $\omega_P^2$),
while the spectrum above $\omega_c$ represents interband transitions.
In the case of pnictides $\omega_c$ is
usually assumed around $2000-3000$
cm$^{-1}$.\cite{qz,yang,wu1,tu,barisic} 

\begin{figure}[t]
\includegraphics[scale=0.38, clip=]{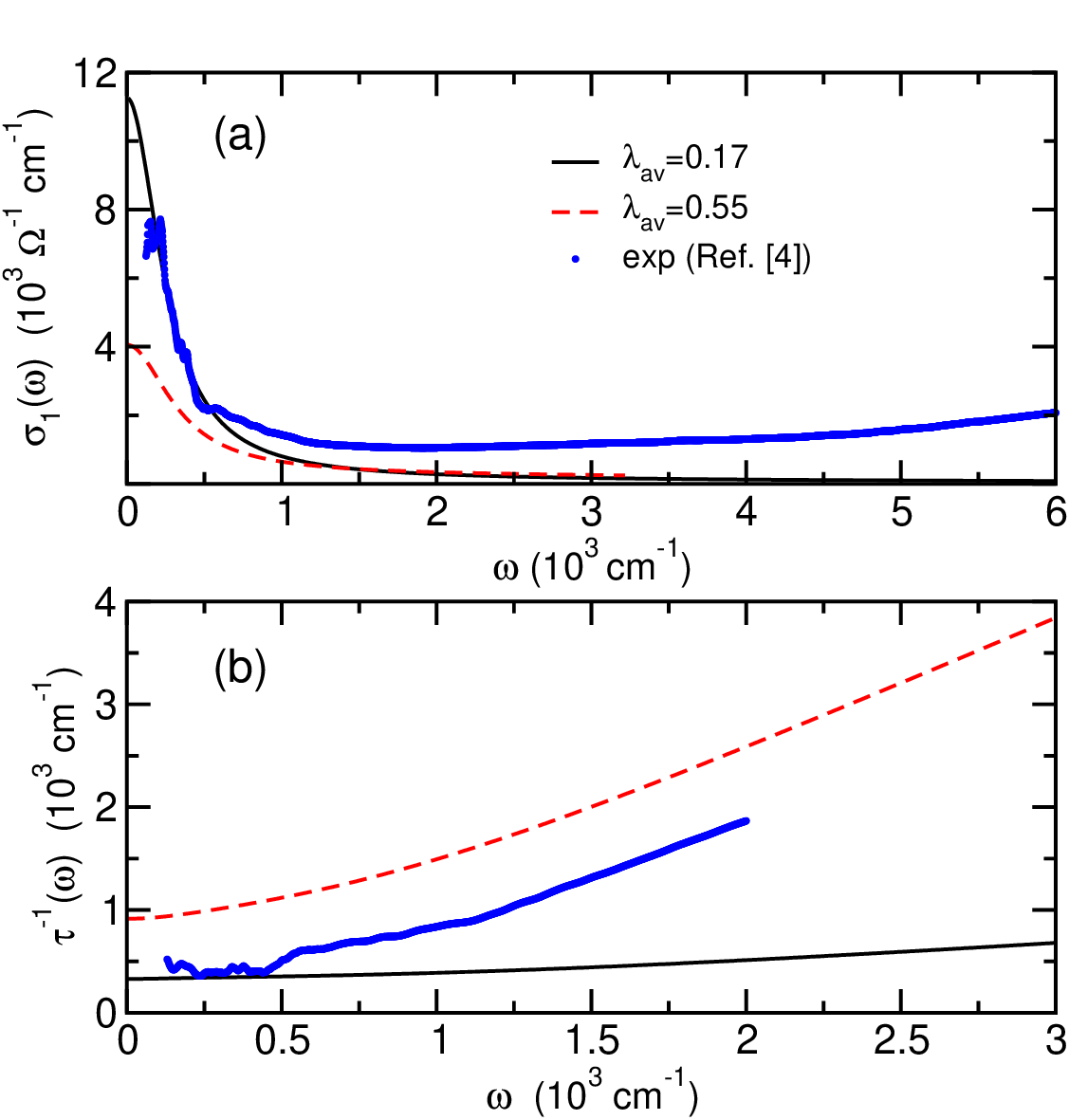}
\caption{(color online) Comparison between experimental data
(filled circles, after Ref. \onlinecite{qz}) and theoretical
calculations of the intraband contribution 
(lines)
for the optical conductivity (panel a) and the
corresponding  $\tau^{-1}(\omega)$ as extracted
by an EDM analysis (panel b). 
Different values of the average
coupling strength are here considered (see text).}
\label{f-sigmal}
\end{figure}

Experimental data for $\sigma_1(\omega)$ and for the resulting
$\tau^{-1}(\omega)$ in LaFePO, from Ref. \onlinecite{qz}, are shown in
Fig. \ref{f-sigmal} as blue symbols.  In particular, the analysis of
$\tau^{-1}(\omega)$ clearly reveals two distinct regimes which are
also visible in $\sigma_1(\omega)$: a low-energy one below $\sim 500$
cm$^{-1}$, where $\tau^{-1}(\omega)$ is roughly constant
($\tau^{-1}(\omega) \approx 400$cm$^{-1}$) and which corresponds to a
very narrow Drude-like contribution in $\sigma_1(\omega)$; and a
higher-energy regime ($\omega > 500$ cm$^{-1}$) where
$\sigma_1(\omega)$ is almost constant and $\tau^{-1}(\omega)$ has a
steady increase, with $\tau^{-1}(\omega) \sim \omega$.
Note that scattering by impurities leads
to a constant $\tau^{-1}(\omega)$, while scattering by retarded interactions is
usually reflected in an increase of $\tau^{-1}(\omega)$ up a saturation regime
at frequencies larger than the typical energy scale
of the interaction.\cite{Shulga} Since phonon frequencies in pnictides
do not exceed $500$ cm$^{-1}$,~\cite{boeri}
one would be tempted to interpret the lack of saturation in
$\tau^{-1}(\omega)$ as due to strong scattering by some other collective
modes,\cite{qz} such as spin fluctuations, that are also
the best candidates for the superconducting pairing.

To investigate the possible role of spin fluctuations on the optical
spectra of LaFePO we calculated the intraband
optical conductivity of the
fully interacting multiband model for LaFePO
already introduced in Refs.~[\onlinecite{ortenzi,sumrule}].
In particular, the charge carriers interact with spin fluctuations modeled in terms of a Millis's
spectrum,\cite{millis_prb92}
\begin{equation}
B_{\rm sf}(\omega)
=
\lambda_{\rm av} 
\frac{\omega\omega_0}{\pi(\omega_0^2+\omega^2)},
\label{millis}
\end{equation}
which emphasizes the survival of bosonic excitations up to large
energies. Here $\omega_0$ is the characteristic energy scale of spin
fluctuations and $\lambda_{\rm av}$ is an average of the dimensionless
coupling constant in each of the four bands included in the
microscopic model.\cite{ortenzi,sumrule} This 
has been estimated to be $\lambda_{\rm av} \approx 0.5$ at $T_c=7$ K ,\cite{ortenzi} to
account for the superconducting,\cite{kamihara} magnetic,\cite{coldea}
and specific heat measurements.\cite{mcqueen,kohama} This coupling is
expected to be reduced at $T=300$ K by roughly a factor 3,
$\lambda_{\rm av} \approx 0.17$, due to the weakening of the
antiferromagnetic correlations, accompanied also by a corresponding
shift at higher energies of  the characteristic energy scale
$\omega_0=60$ meV $\approx 480$ cm$^{-1}$ of the
spin fluctuations.\cite{sumrule,inosov} The complex optical
conductivity is then computed using a finite quasi-particle scattering
rate $\Gamma_0=10$ meV $\approx 81$ cm$^{-1}$ to account for
impurities, and an EDM is applied to these data to extract
$\tau^{-1}(\omega)$, using the experimental plasma frequency
$\omega_P=14900$ cm$^{-1}$.\cite{qz} The results for
$\sigma_1(\omega)$ and $\tau^{-1}(\omega)$ obtained in this way
are shown in Fig.\ \ref{f-sigmal} as solid black lines, in comparison
with the experimental data of Ref. \onlinecite{qz}.  As one can see,
while the low-energy part of $\sigma_1(\omega)$ is well described by
the calculated intraband optical conductivity for $\lambda_{\rm
  av}=0.17$, at $\omega \gtrsim 500$ cm$^{-1}$ the theoretical
$\sigma_1(\omega)$ deviates from the experimental data. This behavior
is reflected also in the corresponding theoretical
$\tau^{-1}(\omega)$, which
increases approximately linearly as $2\lambda_{\rm av}\omega$, i.e
with a slope about four times smaller than the experimental one.
Indeed, within this framework, we are able to reproduce the experimental
linear slope of $\tau^{-1}(\omega)$ at $\omega \sim 1500$ cm$^{-1}$ only by
assuming a total coupling of the order of $\lambda_{\rm av} \approx
0.55$ (red dashed line in Fig. \ref{f-sigmal}).  However such a large
value of $\lambda_{\rm av}$, besides being at odds with the dHvA and
other thermodynamical measurements,\cite{cv,ortenzi,sumrule} would
result in a $\tau^{-1}(\omega=0)$, at $T=300$ K, much higher than the
experimental one. Furthermore, $\sigma_1(\omega)$, turns out to be 
too broad in the low-frequency range with respect to 
experimental data. The present results show thus that in LaFePO it is
not possible to describe the optical data in the range $\omega
\lesssim 2000$ cm$^{-1}$ in terms of pure intraband transitions, even
in the presence of scattering mediated by 
a spin-fluctuation spectrum with a long tail at high energies
as in Eq. (\ref{millis}), and regardless the specific
value of the coupling.

Since a model with intraband transitions alone cannot reproduce the
optical data in pnictides, we investigate the possibility that the 
EDM analysis of the experimental data in LaFePO is
affected by interband transitions having a sizable spectral weight
at energies comparable to the characteristic boson frequencies 
(60 meV $\sim$ 480 cm$^{-1}$ in our case). To elucidate in general the
role of interband transitions we illustrate  
\begin{figure}[t]
\includegraphics[scale=0.38, clip=]{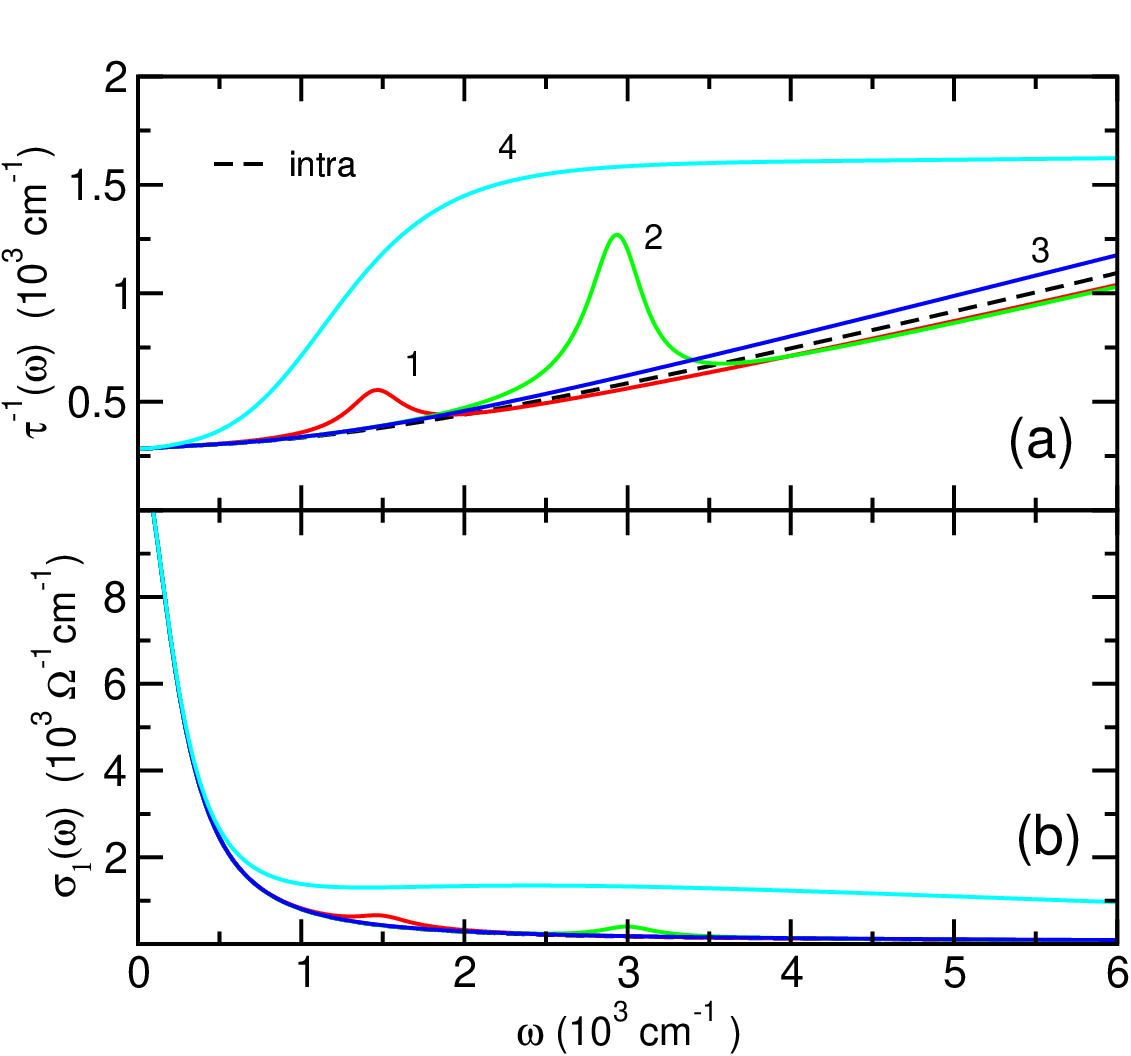}
\caption{(color online) Scattering rate (panel a)  
resulting from the EDM analysis of the optical conductivity $\sigma(\omega)$ from Eq.\
(\ref{lorentz}), and its real part $\sigma_1(\omega)$ 
(panel b). The intraband term is computed here with $\lambda_{\rm av}=0.17$ and
the same microscopic parameters as in Fig.\ \ref{f-sigmal}. The 
Lorentzian parameters are:
 (1) $\omega_L=1500$ cm$^{-1}$, $\gamma_L=400$ cm$^{-1}$,
$S_L=2400$ cm$^{-1}$; (2) $\omega_L=3000$ cm$^{-1}$, $\gamma_L=400$ cm$^{-1}$,
$S_L=2400$ cm$^{-1}$; (3) $\omega_L=3000$ cm$^{-1}$, $\gamma_L=8000$ cm$^{-1}$,
$S_L=2400$ cm$^{-1}$; (4) $\omega_L=3000$ cm$^{-1}$, $\gamma_L=8000$ cm$^{-1}$,
$S_L=24000$ cm$^{-1}$.}
\label{f-tau}
\end{figure}
in Fig.\ \ref{f-tau} a few paradigmatic examples, where we add to the 
intraband term with $\lambda_{\rm av}=0.17$ from Fig. \ref{f-sigmal} (shown here
as dashed black line) a Lorentzian contribution,  with a characteristic energy
$\omega_L$, 
with weight $S_L^2$ and width $\gamma$,
which mimics 
interband transitions:
\begin{equation}
\sigma(\omega)=
\sigma_{\rm intra}(\omega)+
\frac{S_L^2}{4\pi}\frac{\omega}{\omega
 \gamma_L+i(\omega_L^2-\omega^2)}.
\label{lorentz}
\end{equation}
As one can see in Fig.\ \ref{f-tau}a, the effects on $\tau^{-1}(\omega)$
depend crucially on the weight of the Lorentzian and on its
relative
width $\gamma_L/\omega_L$.
For $\gamma_L/\omega_L\ll 1$ (cases 1, 2 in Fig.\ \ref{f-tau}) the Lorentzian
peak can be clearly identified in the real part of the conductivity,
and it also gives a sharp signature in $\tau^{-1}(\omega)$. However, when 
$\gamma_L/\omega_L\simeq 1$ (cases 3, 4) the interband contribution adds
smoothly to the intraband part, see Fig.\ \ref{f-tau}b. This leads in
turn to an overall smooth increase of the
inverse lifetime, that is quantitatively more pronounced when
$S_L^2$ becomes comparable to the spectral weight $\omega_P^2$ of the
intraband part ($\omega_P=14900$ cm$^{-1}$ as in Fig.\ \ref{f-sigmal}). 
As a consequence, in this case the EDM
analysis of the overall conductivity would lead to a
misinterpretation of the effects of the interaction, since the dramatic
increase of $\tau^{-1}(\omega)$ due to the interband transitions could be
interpreted as due to interactions. As we shall argue below, this
could be indeed the case in LaFePO.
\begin{figure}[t]
\includegraphics[scale=0.38, clip=]{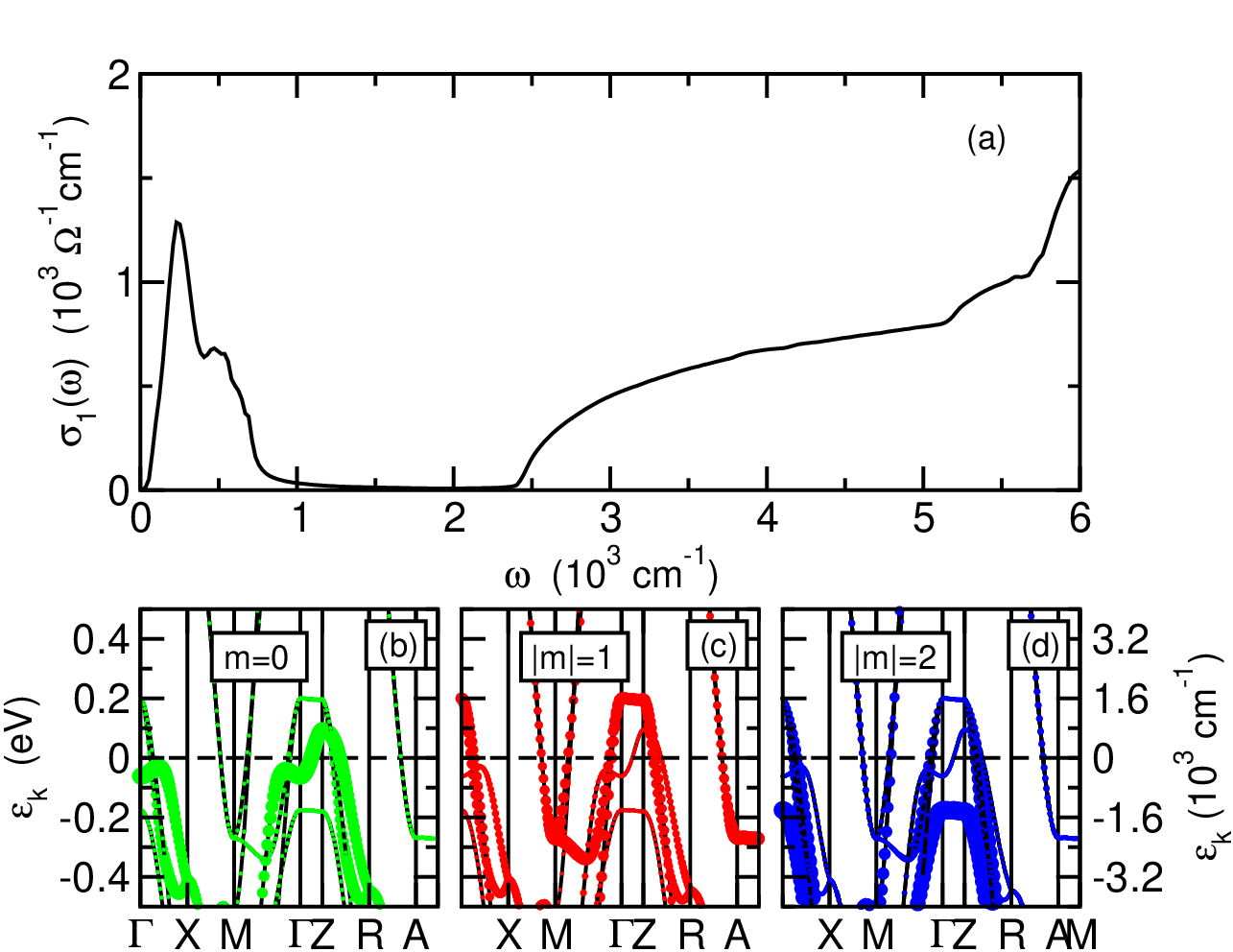}
\caption{(color online) 
(a) Real part of the interband optical conductivity of LaFePO
evaluated in DFT calculations. (b)-(d)
Low-energy band-structure of
LaFePO, where the thickness of the symbols represents the character
$|m|=0,1,2$ of each band.}
\label{f-dft}
\end{figure}

To establish the effective role of interband transitions in LaFePO we
first estimate their strength and position by computing its complex optical conductivity using Density 
Functional Theory (DFT) in the generalized gradient approximation (GGA).~\cite{DFT:PBE}
We employed the full potential linearized augmented planewave (LAPW)
method as implemented in the Wien2k code.~\cite{oka75,wien2k,optics}
The Brillouin zone sampling for the optical calculations was done
using a very dense grid of more than 5000 {\bf k}-points
in the symmetry irreducible wedge.
The complex interband optical conductivity $\sigma(\omega)$
was  evaluated using a broadening $\eta=1$ meV.
We considered experimental values for both the
lattice constants ($a=3.964$ \AA, $c=8.512$ \AA)
and  the internal coordinate $z_P=0.3661$.
The interband contribution to $\sigma_1(\omega)$ is shown in  
panel (a) of Fig.\ \ref{f-dft}. Our results are in 
excellent agreement at all energies with the DFT spectra in literature.\cite{qz} 
For $\omega < 6000$ cm$^{-1}$,
the spectrum shows two main features: a narrow peak at very low energies
($\omega \approx 200-600$ cm$^{-1}$),
and a broad structure starting at $\omega \approx 2500$ cm$^{-1}$.
Note {\em en passant} that,
according to the above discussion regarding
the effects of a Lorentzian interband term illustrated in Fig.\ \ref{f-tau}, 
the broad high-energy feature in Fig.\ \ref{f-dft}a
is expected to have a considerable effect on the inverse lifetime,
while the low-energy peak is expected to produce only modest effects. 

The microscopic origin of the two features in the DFT optical
conductivity is shown in the bottom
panels of Fig.\ \ref{f-dft}, where we decorate 
the bands of LaFePO with partial orbital characters.
In an energy range $\sim 1$ eV around the Fermi level,
the bands have mainly Fe $d$ character;\cite{DFT:Lebegue} optical transitions
are possible for $\Delta m=\pm 1$, {\em i.e.}
between  $m=0$ ($d_{z^2}$) and $|m|=1$
($d_{xz}, d_{yz}$), or between  $|m|=1$ and 
$|m|=2$ ($d_{x^2-y^2}, d_{xy}$).~\cite{optics}
The narrow low-energy peak is mostly due to the transitions
around the $\Gamma$-X and Z-R between the three-dimensional 
$d_{z^2}$ band with $m=0$ 
and the two-dimensional hole bands with character
$d_{xz}, d_{yz}$ ($|m|=1$).
Instead, the higher energy feature mainly involves
transitions along $\Gamma$-Z 
between $|m|=1$ states above the Fermi level
and $|m|=2$ states below it.
While the position and shape of the first peak is very sensitive
to computational details and changes for different materials,
the second, broad feature, which extends up to 16000 cm$^{-1}$,
 is a robust property of the optical spectra of Fe pnictides and
chalchogenides.\cite{haule_prl08,ferber,sanna,charnukha,yin}

It should be remarked however that,
although DFT calculations provide an important
insight into the existence and nature of low-energy
interband transitions we cannot use directly
the DFT data to fit the spectrum of Ref.~\onlinecite{qz}.
In fact, due to the underestimation of electronic correlations,
DFT is known to locate the Fe $d-d$
transitions at larger energy as compared to experiments.\cite{haule_prl08}.
For instance, in  Ref.~\onlinecite{charnukha},
a direct comparison of  experimental
and DFT optical spectra of the Ba 122 compound
has shown that 
the experimental position of the Fe $d$ transitions
is rescaled by a factor of two with respect to DFT calculations.
 
Since the exact scaling factor for LaFePO is not known,
in order to assess quantitatively the role of interband
transitions on the spectra of Ref. \onlinecite{qz},
we model
the  infrared interband
conductivity with 
two Lorentzian
peaks at low energies,
\begin{equation}
\sigma(\omega)=
\sigma_{\rm intra}(\omega)+
\sum_{i=1,2}\frac{S_i^2}{4\pi}\frac{\omega}{\omega
 \gamma_i+i(\omega_i^2-\omega^2)},
\label{lorentz2}
\end{equation}
plus a high-energy dielectric constant $\epsilon_{\rm high}$ to
account for processes at higher energies. 
The intraband part is evaluated for the same microscopic parameters
as in Fig. \ref{f-sigmal}  and for
$\lambda_{\rm av}=0.17$, which
reproduces the correct experimental behavior
of the low-energy Drude-like part of $\sigma_1(\omega)$
for $\omega \lesssim 500$ cm$^{-1}$ (see Fig.\ \ref{f-sigmal}).
The Lorentzian parameters 
are meant instead to model the two 
groups of transitions shown by DFT calculations.
The specific values for the frequency ($\omega_i$), width
($\gamma_i$) and weight ($S_i$) of the two Lorentzian contributions,
chosen to fit
the experimental
data for $\sigma_1(\omega)$ in the range $\omega \in [500:2000]$ cm$^{-1}$,
are reported in Table \ref{t-model}.

\begin{table}[t]
\begin{center}
\begin{tabular}{|c|c|c|c|c|c|c|}
\hline \hline
$S_1$ & $\omega_1$ & $\gamma_1$ &
$S_2$ & $\omega_2$ & $\gamma_2$ & $\epsilon_{\rm high}$
\\
\hline 
$2300$ & $750$ &  $40$ &
$33500$ & $4500$ & $20200$ & $15$
\\
\hline
\end{tabular}
\end{center}
\caption{Interband parameters of the model (\ref{lorentz2})
expressed in units of cm$^{-1}$ ($S_i, \omega_i, \gamma_i$) and in
dimensionless
units ($\epsilon_{\rm high}$).}
\label{t-model}
\end{table}

\begin{figure}[t]
\includegraphics[scale=0.38, clip=]{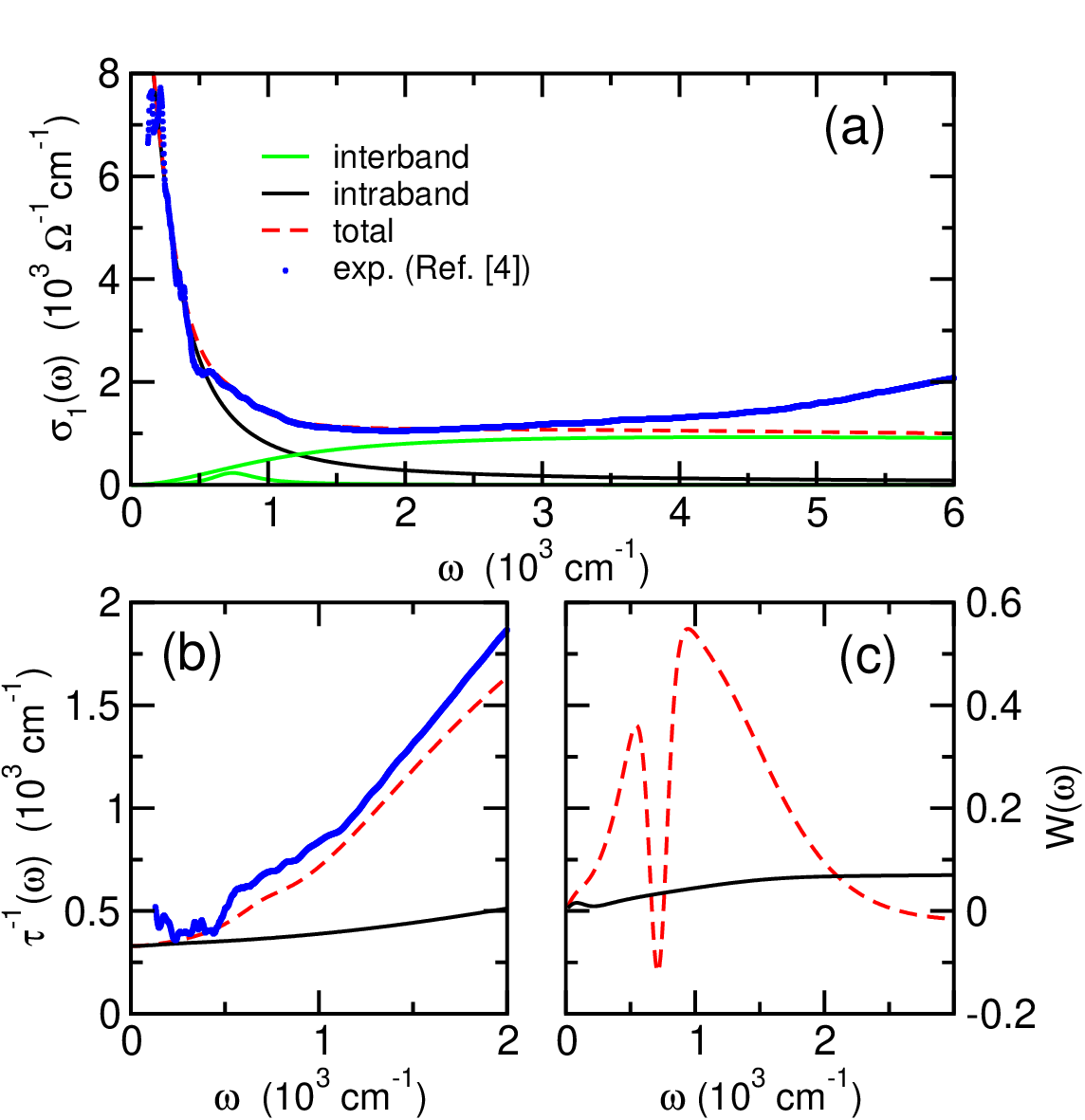}
\caption{(color online) (a) Real part of the optical conductivity
  $\sigma_1(\omega)$ computed using Eq. (\ref{lorentz2}) (red dashed 
  line, labeled as ``total'' in the caption), along with the intraband
  part (solid black line) and the interband part (solid green
  line). The symbols represent the experimental data of
  Ref. \onlinecite{qz}. (b) $\tau^{-1}(\omega)$ extracted from the
extended-Drude-model using the total conductivity (dashed red line) or the
  intraband part only (solid black line), along with the experimental
  data (symbols)  (c) Approximate
  electron-boson Eliashberg function $W(\omega)$ extracted from
  $\tau^{-1}(\omega)$ with the second-derivative method (see text)
  using either the total conductivity (dashed red line) or the
  intraband part only (solid black line).}
\label{f-our}
\end{figure}

The result for the total optical conductivity is shown in black line in
Fig. \ref{f-our}a, where one can see the very good agreement with the
experimental data from Ref. \onlinecite{qz} in the whole range $\omega
\in [200:3000]$ cm$^{-1}$.
The effect of the interband transitions on
the EDM analysis is elucidated in Fig.\ \ref{f-our}b
where we compare $\tau^{-1}(\omega)$ calculated by
using both the total optical response (dashed line) and the interacting
intraband part only (solid line).  As one can see in Fig. \ref{f-our}b,
the presence of interband transitions remarkably affects
the magnitude and the frequency dependence of the effective
EDM functions of LaFePO at very low energies ($\omega \gtrsim
500$ cm$^{-1}$), in analogy with the case (4) of Fig.\ \ref{f-tau}
discussed above. 
In particular,
the inverse lifetime $\tau^{-1}(\omega)$ acquires a frequency
dependence that is much stronger than the one of
the intraband contribution alone, with a linear slope
which is $\approx 4$ times larger 
than the one arising from the actual intraband contribution.

In the spirit of the EDM, the frequency
dependence of $\tau^{-1}(\omega)$ is often used to extract the
underlying spectrum $\alpha^2F(\omega)$ of the retarded interaction
supposed to generate it.
A crude approximation to obtain the interaction spectrum is the second
derivative method, in which $\alpha^2F(\omega)$ is approximated by
$W(\omega)$, where
$W(\omega)=(1/2\pi)d^2
[\omega\tau^{-1}(\omega)/d\omega^2]$.\cite{norman,carbotte2}
This
method is usually sufficient to capture the order of magnitude of the
spectrum, although not its fine details.  The $W(\omega)$
corresponding to the total and to the pure intraband $\tau^{-1}(\omega)$ are shown
as dashed red and solid black lines, respectively, in
Fig. \ref{f-our}c, showing that the spectrum extracted from the
total $\tau^{-1}(\omega)$ is significantly larger than the
other one.  Indeed, the ``effective'' coupling $\lambda_{\rm
  eff}=2\int_0^{\omega_c} d\omega W(\omega)/\omega$, with
$\omega_c=3000$ cm$^{-1}$, obtained from the intraband term
results
$\lambda_{\rm eff}=0.27$, larger but of the same order of magnitude 
as the microscopic value used here, $\lambda_{\rm av}=0.17$,\cite{notel}
whereas the effective coupling estimated using the total
$\tau^{-1}(\omega)$ would give $\lambda_{\rm eff}=1.31$, significantly
larger 
than $\lambda_{\rm av}$. This observation can thus explain why a direct EDM
analysis of optical data, without a proper subtraction of the interband
processes, gives much larger values of the coupling~\cite{yang,wu1} than what
estimated by other probes.~\cite{ortenzi,mcqueen,kohama}

In summary, we have shown that the EDM analysis in pnictide systems
can be strongly biased by the presence of significant interband
contributions in the mid-infrared region of the optical spectrum. 
In particular, the strong increase of the inverse lifetime $\tau^{-1}(\omega)$,
observed in several compounds,\cite{qz,yang,wu1,tu,barisic} cannot be
straightforwardly attributed to strong-coupling effects, as it would
be the case for optical spectra dominated by intraband interactions. 
Even though we focused explicitly on LaFePO, we expect similar effects to
hold in other classes of pnictides as well, due to the very similar
band structure. In particular, the presence of low-energy interband
transitions can provide an alternative explanation for the optical
spectra of 122 compounds, where the flat region of the
$\sigma_1(\omega)$ has been usually fitted with a Drude component having an
unrealistically large scattering rate.\cite{tu,wu4}

\acknowledgements
We thank D. Wu for many useful discussions,
and the authors of Ref. \onlinecite{qz}
for providing us with the experimental data for comparison.
L. Boeri. and L. O. would like to thank O. Dolgov, A. Boris and
A. Charnukha for useful discussions, and for suggesting us
Refs. \onlinecite{Shulga,physica_c}.
L. O. acknowledges funding from DFG SPP 1458, proj. Bo-3536/1.
L.B. and E.C. acknowledge partial funding from the Italian MIUR
under the project  PRIN 2008XWLWF9. 

While completing the present manuscript, we became aware of a work, 
which discusses experimentally
the role of low-energy interband transitions on the optical scattering 
of K-doped BaFe$_2$As$_2$\cite{boris2}.


\begin{thebibliography}{99}
\bibitem{notel}
For the pure intraband case,
moderate deviation from $\lambda_{\rm eff}$ and $\lambda_{\rm av}$
are expected to arise due to the finite temperature effects
and due to the failure, in multiband systems, of the Allen approximation used
to define the second derivative method.
\end{thebibliography}
\end{document}